\documentclass[acus,
               biblatex,
              ]{jacow}

\makeatletter
\ifboolexpr{bool{xetex}}                 
 {\renewcommand{\Gin@extensions}{.pdf,%
                    .png,.jpg,.bmp,.pict,.tif,.psd,.mac,.sga,.tga,.gif,%
                    .eps,.ps,%
                    }}{}
\makeatother

\ifboolexpr{bool{xetex} or bool{luatex}} 
 {}                                      
 {\usepackage[utf8]{inputenc}}           

\usepackage[USenglish]{babel}

\ifboolexpr{bool{jacowbiblatex}}
 {%
  \addbibresource{RFQ_Injection.bib}
 }{}

\usepackage{xspace}
\newcommand{\degree}{\ensuremath{^\circ}\xspace}

\newcommand{\htp}{H$_2^+$\xspace}
\newcommand{\nuebar}{$\bar{\nu}_e$\xspace}

\begin{document}

\title{An RFQ Direct Injection Scheme for the IsoDAR High Intensity \htp 
Cyclotron\thanks{Work supported by grant NSF-PHY-1148134 and the 
Massachusetts Institute of Technology.}}

\author{D. Winklehner, MIT, Cambridge, MA, USA\\
        R. Hamm, R{\&}M Technical Enterprises, Pleasanton, CA, USA\\
        J. Alonso, J. Conrad, MIT, Cambridge, MA, USA\\
        }

\maketitle

\begin{abstract}
IsoDAR is a novel experiment designed to measure neutrino oscillations through
\nuebar disappearance, thus providing a definitive search for sterile 
neutrinos. In order to generate the necessary anti-neutrino flux, a high 
intensity primary proton beam is needed. In IsoDAR, \htp is accelerated and 
is stripped into protons just before the target, to overcome space charge 
issues at injection. As part of the design, we have refined an old proposal to 
use an RFQ to axially inject bunched \htp ions into the driver cyclotron. This 
method has several advantages over a classical low energy beam transport 
(LEBT) design: (1) The bunching efficiency is higher than for the previously 
considered two-gap buncher and thus the overall injection efficiency is higher. 
This relaxes the constraints on the \htp current required from the ion source. 
(2) The overall length of the LEBT can be reduced. (3) The RFQ can also 
accelerate the ions. This enables the ion source platform high voltage to be 
reduced from 70 kV to 30 kV, making underground installation easier. We are 
presenting the preliminary RFQ design parameters and first beam dynamics 
simulations from the ion source to the spiral inflector entrance.
\end{abstract}

\section{Introduction}

\begin{figure}[!t]
	\centering
		\includegraphics[width=1.00\columnwidth]
        {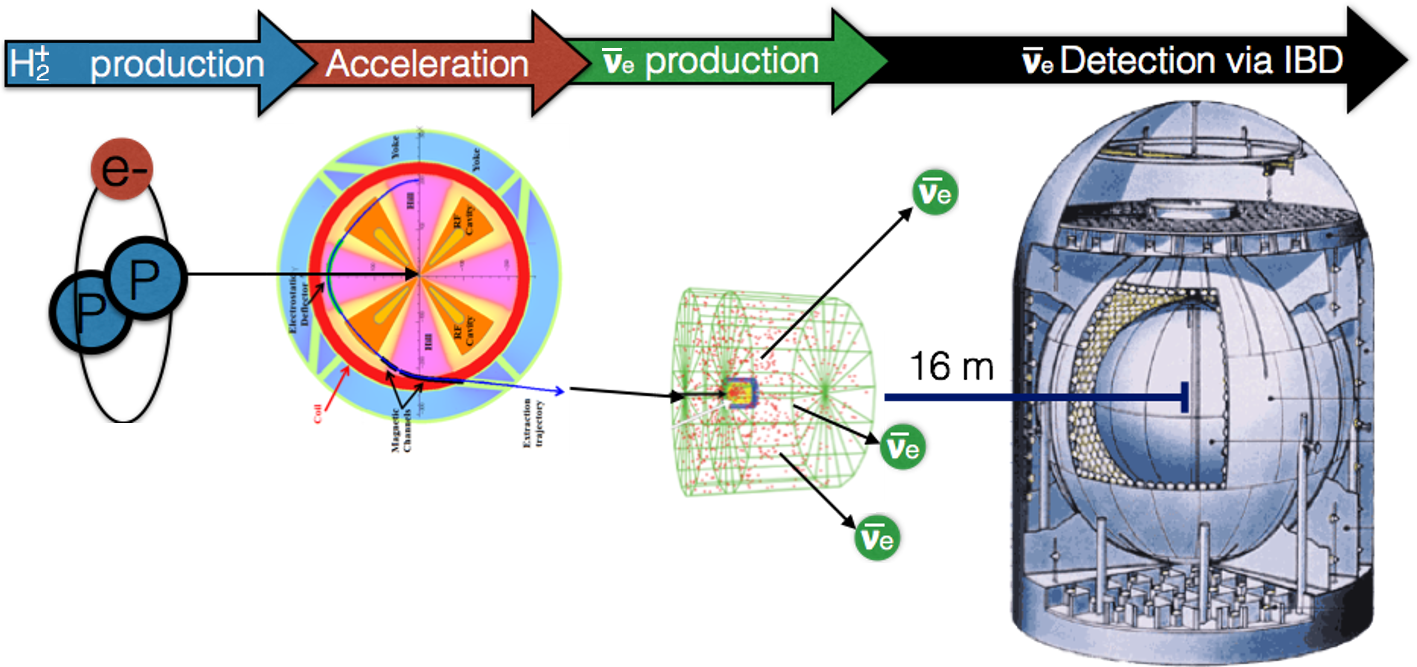}
	    \caption{Cartoon of the IsoDAR experiment. Detector 
	             image courtesy of the KamLAND collaboration.}
	\label{fig:isodar}
\end{figure}

In the IsoDAR (Isotope Decay-At-Rest) experiment \cite{adelmann:isodar}, \htp 
ions will be delivered by a high performance ion source, accelerated to 60 
MeV/amu in a compact cyclotron, and then stripped to protons and dumped on a 
beryllium target surrounded by a lithium sleeve. Neutrons generated by the 
protons hitting $^9$Be will be captured on $^7$Li, and the resulting $^8$Li will then 
beta decay-at-rest, producing a very pure \nuebar beam. This accelerator and 
target system will be placed close (16 m from the center) to an existing 
neutrino detector (e.g. KamLAND) to measure \nuebar disappearance from neutrino 
oscillations via inverse beta decay. 
This process is depicted in Figure \ref{fig:isodar}.
The short baseline will allow tracing out more than a full period of the 
oscillation wave inside the detector, 
thus presenting a definitive search for proposed sterile neutrinos that may 
participate in the oscillation but not in the weak interaction. 
In order to get definitive results over the course of a few years, a high 
neutrino flux is necessary; hence a very high primary proton beam is desired. 

In the summer months of 2013 and 2014, tests of high intensity \htp production, 
transport, and injection into a cyclotron were conducted at Best Cyclotron Systems,
Inc. (BCS) in Vancouver, Canada. The ion source, borrowed from INFN Catania in Italy, 
was the "versatile ion source" (VIS), an off-resonance 2.45 GHz ECR ion source. From
the VIS, it was possible to extract 12 mA of \htp. The main objective was to test
the extent to which space charge will be a limiting factor in the low energy beam 
transport and the injection into the cyclotron through a spiral inflector. It was 
possible to show that the transmission through the spiral inflector at beam currents 
on the order of 10 mA was $\approx 95\%$. The typical acceptance of an unbunched 
beam into the cyclotron RF bucket is on the order of 5\%, and with a 2-gap 
multiharmonic buncher, this acceptance can be increased to 10-20\%. Considering 
the present performance of the VIS and suggested improvements to the source, the 
nominal 5 mA of \htp extracted from the cyclotron is achievable, but pushing the 
limits. We are exploring two avenues to improve the situation: (1) We are 
constructing a new ion source (multicusp) dedicated to the production of \htp, and 
(2) we are investigating the use of an RFQ to replace the LEBT.
The RFQ is a linear accelerator that can focus, bunch, and accelerate a continuous 
beam of charged particles at low energies with high bunching and transmission 
efficiencies. RFQs are very attractive for low energy ion 
accelerators, e.g. for applications with high current beams or in combination
with sources such as an ECR, because the source can be close to ground potential 
and is easy to operate and to service. Because the basic RFQ concept can be 
implemented over a wide range of frequencies, voltages, and physical dimensions, 
it is an ideal structure to use for bunching an intense ion beam for injection 
into a cyclotron and has already been used in several cyclotron systems for 
radial injection of input beams \cite{schempp:rfq1}.  
It can also be used for axial injection at much lower energies as first proposed 
in 1981 \cite{hamm:rfq1}. However, to date, direct axial injection into a compact 
cyclotron using an RFQ has not yet been realized.
For high acceptance into the cyclotron, the injected beam should be bunched at 
the cyclotron frequency (33.2 MHz at 4th harmonic), resulting in the use of the 
four-rod RFQ structure \cite{schempp:rfq2}.

\raggedend
\section{RFQ Design}

\begin{figure}[!t]
	\centering
		\includegraphics[width=0.7\columnwidth]
        {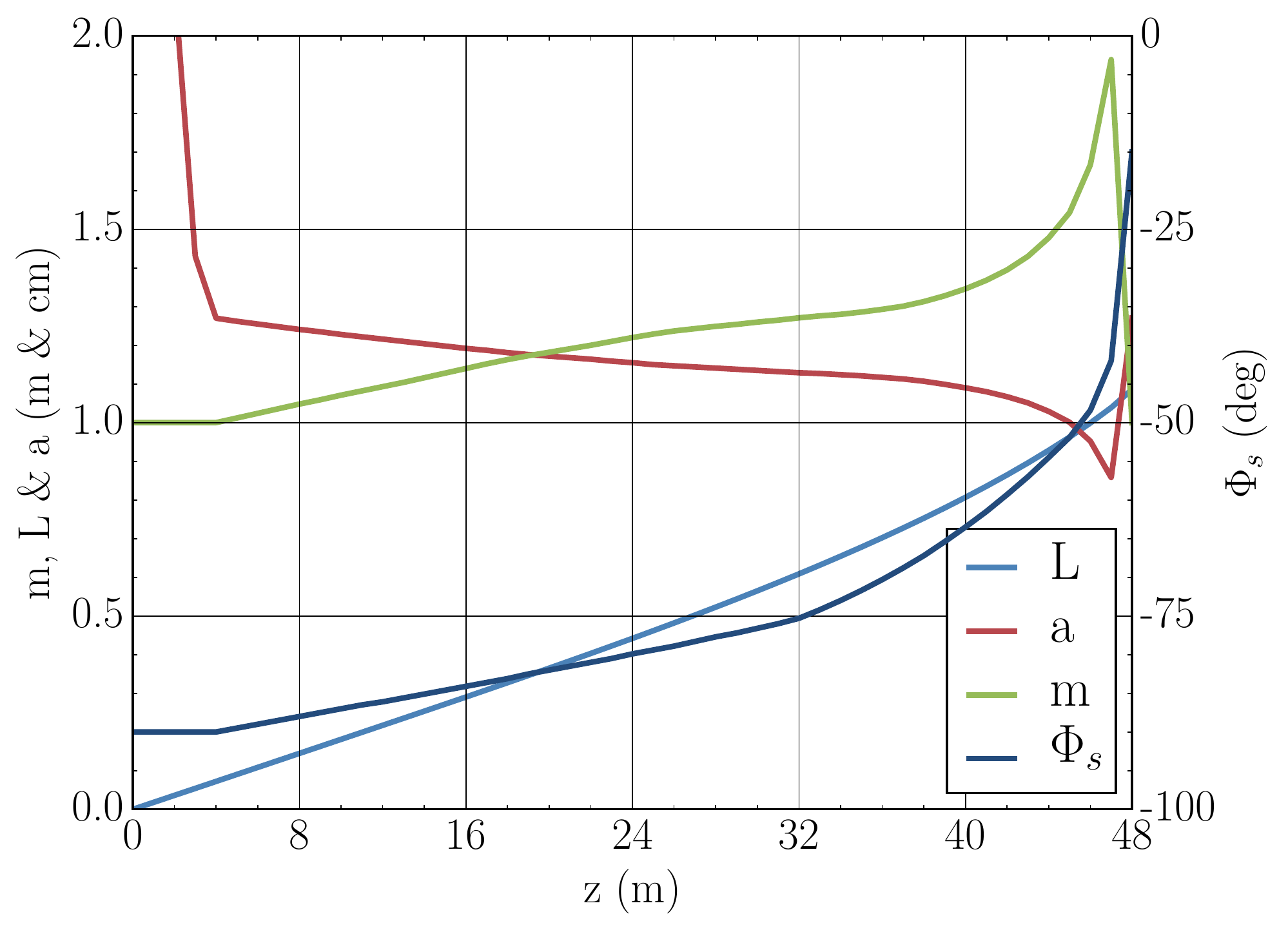}
	\caption{RFQ-Linac design parameters for IsoDAR direct injection.}
	\label{fig:params}
\end{figure}

Such an RFQ has been designed for axial injection into the IsoDAR cyclotron. The
design parameters of this injector RFQ are listed in Table \ref{tab:params}.
This four-rod structure accelerates and bunches \htp ions extracted from an ECR
ion source for injection into the spiral inflector in the cyclotron. To minimize the 
RFQ's radial dimension, distributed spiral inductor resonators would be used.

\begin{table}[!b]
    \centering
    \caption{IsoDAR RFQ-Linac Injector Parameters}
    \begin{tabular}{ll}
        \toprule
        \textbf{Parameter} & \textbf{Value} \\
        \midrule
            Operating frequency	& 33.2 MHz \\
            Injection energy & 15 keV \\
            Final beam energy & 80 keV \\
            Design input current & 10 mA \\
            Current limit & 22 mA \\
            Transmission at 10 mA & 99\% \\
            Inp. trans. emittance (6-rms, norm.) & 0.5 $\pi$-mm-mrad \\
            Nominal vane voltage & 43 kV \\
            Bore radius & 1.27 cm \\
            Maximum vane modulation & 1.94 \\
            Structure length & 1.09 m \\
            Peak RF field surface gradient & 4.66 MV/m \\
            Structure RF power & 0.82 kW \\
            Beam power & 0.64 kW \\
            Total input RF power & 1.46 kW \\
        \bottomrule
    \end{tabular}
    \label{tab:params}
\end{table}

\section{Parmteq Simulations of RFQ}

\begin{figure}[!t]
	\centering
		\includegraphics[width=0.8\columnwidth]
		                {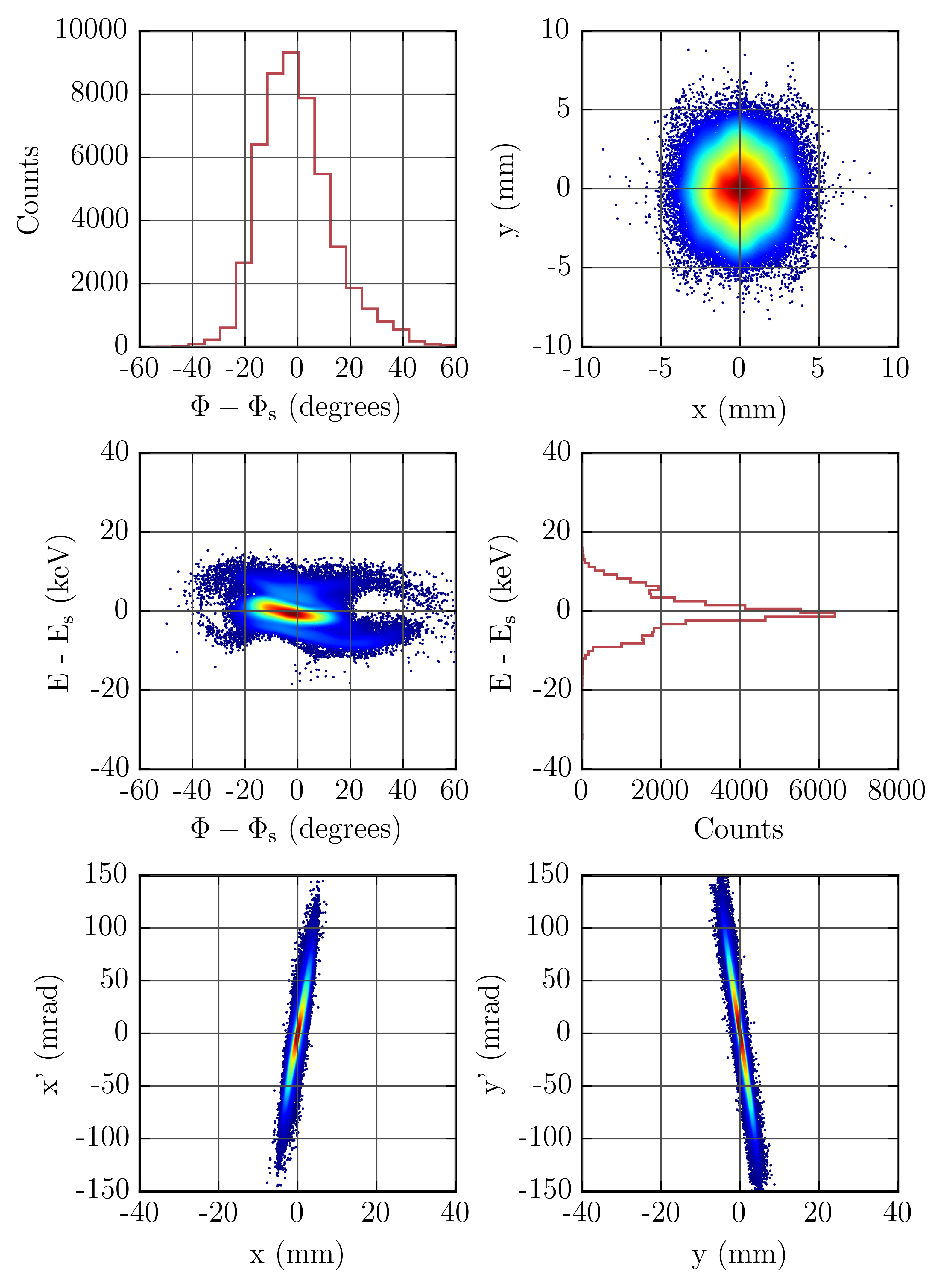}
	\caption{Phase Spaces of the RFQ output beam.}
	\label{fig:phasespace1}
\end{figure}

This RFQ, which is used primarily for bunching the beam, has only 48 cells. 
The design parameters of the RFQ vanes are shown as a function of cell number in 
Figure \ref{fig:params}. 
Figure \ref{fig:phasespace1} shows the calculated phase space projections at the 
output of the RFQ using 50,000 macroparticles in the beam dynamics code PARMTEQ 
\cite{crandall:parmteq1}. The total phase width of the beam is ~ 90\degree, but 
60\% of the beam is within $\pm10\degree$. The energy spread of the beam at 80 
keV has a FWHM value of 5 keV but has a “halo” that extends out to $\pm1.5\%$. 
Figure \ref{fig:phasespace1} shows the output phase spaces in all three planes. 
As seen in the figure, the beam in the transverse planes exiting the RFQ is 
almost circular with a diameter of $\approx$6mm. The transmission of protons is 0\%
showing very good \htp beam purity

\section{Ion Source and Matching to RFQ}

\begin{figure*}[t]
	\centering
	\includegraphics*[width=0.8\textwidth]
                     {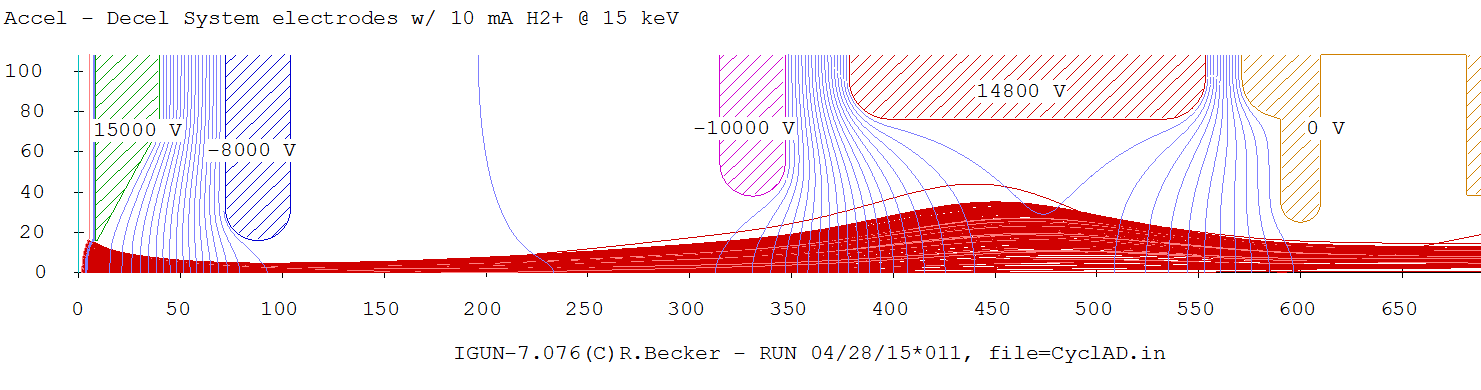}
	\caption{IGUN simulation of beam extraction and matching to RFQ.
	         \label{fig:igun}}
\end{figure*}

Two types of ion sources are currently being considered for IsoDAR: A new 2.45 
GHz off-resonance ion source comparable to the VIS and a multicusp ion source with 
a short plasma chamber (a prototype is currently being built at MIT). Both are plasma
type ion sources, and the extraction can be simulated with the well known code IGUN 
\cite{becker:igun1}. 
For the IsoDAR cyclotron, the CW beam from the ion source can be extracted
in an accel–decel extraction system, and focused into the RFQ using a segmented
einzel lens. As an example, the optics for a 10 mA beam coming from a VIS-type source, calculated with IGUN, is shown in Figure \ref{fig:igun}.

\section{Cyclotron Injection Simulations}

\begin{figure}[b]
	\centering
	\includegraphics[height=0.20\textheight]
                    {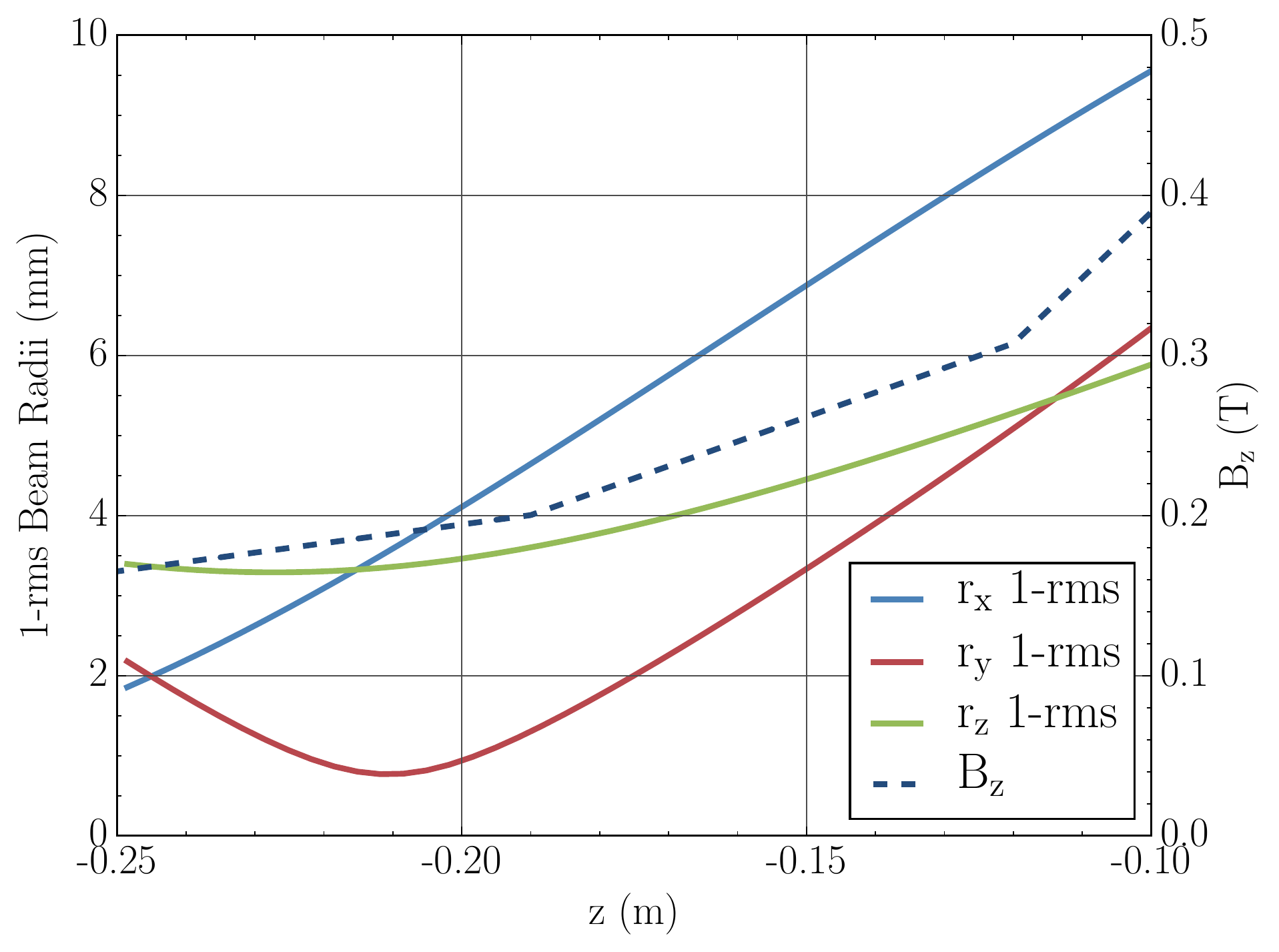}
	\caption{The 1-rms bunch radii for the beam traveling from the exit of the 
	         RFQ to the entrance of the spiral inflector. The residual axial 
	         magnetic field of the cyclotron is plotted on the secondary axis.
	         \label{fig:plug_envelopes}}
\end{figure}

In order to verify the feasibility of the RFQ injection scheme, careful
investigation of the matching of the RFQ exit to the cyclotron is necessary.
The particle-in-cell (PIC) code OPAL \cite{adelmann:opal} was recently updated
with routines to incorporate complicated electrode geometries as boundary 
conditions during the calculation of self-fields and for the termination of 
particles. In addition, OPAL-CYC, the subset of OPAL tailored to cyclotron 
simulations, was upgraded to manage axial injection through a spiral inflector.
The first step in this simulation study was to transport the particles exiting 
the RFQ to the entrance of the spiral inflector in order to see how much of an 
effect the residual magnetic field in the cyclotron plug would have. The 
resulting 1-rms beam radii as well as $\mathrm{B}_\mathrm{z}$(z) on-axis 
are shown in Figure \ref{fig:plug_envelopes}. It becomes immediately clear that 
even with the small focusing effect of the magentic field,
the beam size increases significantly over the distance of $\approx 15$cm. 
Transporting the beam through the spiral inflector, about 80\% of the beam is lost 
before encountering the first acceleration gap.
It is possible to change the divergence and size of the beam at the exit of the 
RFQ by shortening the last cell. However, tests with different beams exiting the 
RFQ yielded similarly unacceptable results.
In order to improve this behaviour in the next design iteration, an additional 
focusing element and a one gap buncher will be placed inside the plug (between 
RFQ exit and spiral inflector entrance) to compensate for the divergence and 
energy spread of the bunch.

\section{Conclusion}

The IsoDAR proton driver aims to deliver 10 mA of protons to a high power 
beryllium target to produce a pure \nuebar beam for a conclusive search for
the existence of sterile neutrinos. \htp was chosen as primary ion to overcome
space-charge issues. First tests with a 2.45 GHz ECR ion source (VIS) and a 
test beam line showed that achieving the necessary injection efficiencies to
ultimately extract 5 mA of \htp from the cyclotron is possible, but pushing 
the limits of classic LEBT design. 
An RFQ closely coupled to the ion source 
upstream and re-entrant to the cyclotron iron yoke can overcome the 
limitations experienced in the preliminary tests. 
In a design study, a 1.09 m 
long RFQ for 33.2 MHz operation was simulated using Parmteq. The results are 
very promising, showing an RFQ output emittance of 1.4 $\pi$-mm-mrad (4-rms, 
normalized). This is similar to the original LEBT design, but with  
transport efficiency increased to 99\% and with 63\% of the beam within $\pm 10 \degree$
RF phase angle and within $\pm 2$ keV energy spread at 80 keV beam energy. 
In addition the system is much more compact, making underground installation easier.
Preliminary injection simulations using the OPAL code suggest that the increase
in beam size from the exit of the RFQ to the entrance of the spiral inflector 
is inhibitively large and thus, the next design iteration will incorporate a 
focusing element and a re-buncher between RFQ and spiral 
inflector.

\section{Acknowledgement}

This work was supported by grant NSF-PHY-1148134 and MIT seed funding 
(PI: Prof. J. Conrad). The authors are thankful to the AMAS group at PSI for 
their support with the OPAL code and Prof. R. Becker for help with IGUN.

%
%

\printbibliography

\end{document}